\documentclass[slac_one]{revtex4}
\usepackage{graphicx}
\usepackage{fancyhdr}
\begin{document}

\title{Final state interaction effects on the $ \eta_{b}\to J/\psi J/\psi $ decay%
\footnote{To the memory of Giuseppe (Beppe) Nardulli}%
} 

%
\author{Pietro Santorelli}
\affiliation{Universit\`a  di Napoli ``Federico II'' \& INFN, Napoli, Italy}

\begin{abstract}
We study the effects of final state interactions on the
$\eta_b\to J/\psi J/\psi$ decay. In particular, we discuss the effects
of the annihilation of $\eta_b$ into two charmed meson and their
rescattering into $J/\psi J/\psi$. We find that the inclusion of this
contribution may enhance the short-distance branching ratio up to
about 2 orders of magnitude.
\end{abstract}

\maketitle

\thispagestyle{fancy}


Large efforts have been invested during the past thirty years to
look for $\eta_{b}$ but the evidence of its existence emerged
very recently thanks to the \textsc{Babar} collaboration \cite{Aubert:2008vj}.
In \cite{Aubert:2008vj} is reported the
first unambiguous evidence of $\eta_{b}$, with a $10\,\sigma$
significance, through the hindered magnetic dipole transition
process $\Upsilon(3S)\to \eta_b\gamma$. The mass of $\eta_{b}$ is also
measured to be  $m_{\eta_{b}}\,=\,9388.9^{+3.1}_{-2.3}({\rm stat})\pm2.7({\rm syst})$ MeV.
Apart from its mass and the branching ratio of the $\Upsilon(3S)\to \eta_b\gamma$, almost nothing is known
regarding the decay pattern of $\eta_{b}$~\cite{Brambilla:2004wf}.
However, rough estimate of the branching ratios of some exclusive
two and three-bodies hadronic decays can be found in~\cite{Jia:2006rx}.\\
Some {\it golden} modes have been proposed to observe $\eta_b$, such
as $\eta_b\to J/\psi J/\psi$~\cite{Braaten:2000cm} and $\eta_b \to
J/\psi\gamma$~\cite{Hao:2006nf,Gao:2007fv}.  Despite very clean
signature due to the $J/\psi$ in final state, these decay modes are
estimated to have rather suppressed branching ratios. Regarding the
$\eta_b\to J/\psi J/\psi$ decay mode, the original estimate~\cite{Braaten:2000cm},
which was compatible with the discovery of $\eta_b$ in Tevatron Run I, has been
reconsidered~\cite{Maltoni:2004hv,Jia:2006rx}. In particular, an explicit
NRQCD calculation gives ${\cal B}r[\eta_b\to J/\psi J/\psi]  = (0.5 \div 6.6) \times 10^{-8}$~\cite{Jia:2006rx}%
\footnote{See also very recent calculation in NRQCD at NLO in $\alpha_{s}$
${\cal B}r[\eta_b\to J/\psi J/\psi] = (2.1 \div 18.9) \times 10^{-8}$~\cite{Gong:2008ue}.}
too small to be observed also in Tevatron Run II.

An interesting decay channel to observe $\eta_b$, $\eta_b\to D^{(\ast)} \overline{D^\ast}$,
has been proposed in \cite{Maltoni:2004hv} where the range
$10^{-3}<{\cal B}r [\eta_b\to D \overline{D^\ast}]<10^{-2}$
and ${\cal B}r [\eta_b\to D^\ast \overline{D^\ast}]\approx 0$ were predicted.
On the other hand, in Ref.~\cite{Jia:2006rx}, by doing reasonable physical considerations, the author obtained
${\cal B}r [\eta_b\to D \overline{D^\ast}] \sim  10^{-5}\, $ and
${\cal B}r [\eta_b\to D^\ast \overline{D^\ast}] \sim 10^{-8}$
which are at odds with the ones obtained in \cite{Maltoni:2004hv}.

\begin{figure}[]
  \includegraphics[width=6.5truecm]{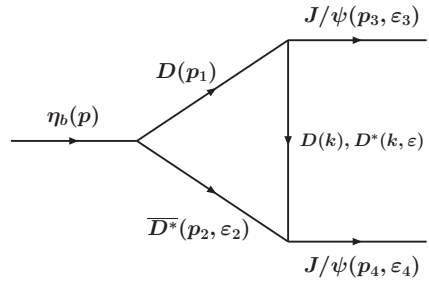}\\
  \caption{Long-distance $t-$channel rescattering contributions to $\eta_b \to J/\psi\ J/\psi $.}
  \label{f:triangle}
\end{figure}

In~\cite{Santorelli:2007xg} we assumed that the long distance contribution to the final state made of two $J/\psi$
is dominated by the $D \overline{D^\ast}$ state and the subsequent rescattering of it into two $J/\psi$ with a charmed
meson in the $t-$channel  as is shown in figure~\ref{f:triangle}.
The branching ratio of $\eta_b\to D \overline{D^\ast}$ is poorly known at present. However, as we already said there are
two theoretical determinations we will use in considering the contribution to the $\eta_b\to J/\psi\ J/\psi$.
Moreover, we will neglect the contribution coming from the annihilation of the $\eta_b$ to $D^\ast \overline{D^\ast}$,
in agreement with the results in \cite{Jia:2006rx,Maltoni:2004hv}.

The dominance of $D \overline{D^\ast}$ intermediate state is a consequence
of the large coupling of $D^{(\ast)} \overline{ D^{(\ast)}}$ to $J/\psi$
as a result of quark models and QCD Sum Rules calculations.

The full amplitude which takes into account the short distance part and the contribution coming from the
evaluation of the graphs in figure \ref{f:triangle} can be written as
\begin{eqnarray}\label{e:fullA}
\mathcal{A}_f(\eta_b(p) \to J/\psi(p_3, \varepsilon_3)\ J/\psi(p_4, \varepsilon_4)) & = &
\imath\, \frac{ g_{\eta _b JJ }}{m_{\eta_b}}\, \varepsilon_{\alpha\beta\gamma\delta}
p_3^\alpha p_4^\beta\epsilon_3^{\ast \gamma} \epsilon_4^{\ast \delta}
\left[1+ \
\frac{g_{\eta _b DD^* }}{g_{\eta _b JJ }} \left(\frac{}{} \imath\ A_{LD} + D_{LD}\right)\right]\,,
\end{eqnarray}
where $A_{LD}$ and $D_{LD}$ represent the absorbitive and the dispersive part
of the graphs in figure \ref{f:triangle}, respectively. For details about
the calculation of the previous quantities we refer to \cite{Santorelli:2007xg,Santorelli:2007vm}.
The coupling $g_{\eta _b JJ }$ is obtained by using the results in \cite{Jia:2006rx}
while $g_{\eta _b DD^\ast}$ from the estimate of the ${\cal B}r[\eta_b \to D \overline{D^\ast}]$ and so
\begin{equation}
{\frac{ g_{\eta _b DD^* }}{g_{\eta _b JJ }} } \,\,
\left\{
\begin{tabular}{clll}
= & $ 1 $        & \, for  $\,  \mathcal{B}r[\eta_{b} \to D \bar{D^{\ast}}] \approx 10^{-5}$ & \cite{Jia:2006rx} \\
&  & & \\
$\in $ & $[11,35]$ & \, for $\,  10^{-3} \leq \mathcal{B}r[\eta_b \to D \overline{D^\ast}] \leq 10^{-2}$ & \cite{Maltoni:2004hv}.\\
\end{tabular}
\right .
\end{equation}
The numerical values of the on-shell strong couplings
$g_{JD D}, g_{JDD^\ast}$ and $g_{JD^\ast D^\ast}$%
\footnote{We use dimensionless strong coupling constants in all cases. In particular we use the ratio
$g_{JDD^\ast}/m_{J/\psi}$ instead of the dimensional $G_{JDD^\ast}$ (GeV$^{-1}$) usually found in literature.}
are taken from QCD Sum Rules \cite{QCDSR}, from the Constituent Quark Meson model \cite{Deandrea:2003pv}
and from relativistic quark model \cite{RQM} findings which are compatible each other.
We used ($g_{JD D},g_{JDD^\ast},g_{JD^\ast D^\ast}$) = $(6, 12, 6)$.
To take into account the off-shellness of the exchanged $D^{(*)}$ mesons
in figure \ref{f:triangle} we have introduced the $t-$dependance of these couplings by means of the function
\begin{equation}
F(t)  = \frac{\Lambda^2 - m_{D^{(\ast)}}^2 }{\Lambda^2-t}\,.
\label{e:formfactor}
\end{equation}
No first-principles calculation of  $\Lambda$ exists, so,  following
the authors of \cite{Cheng:2004ru}, we write
$\Lambda = m_R + \alpha \Lambda_{QCD} $,
where $m_R$ is the mass of the exchanged particle ($D$ or $D^\ast$),
$\Lambda_{QCD} = 220\ MeV$ and  $\alpha \in [0.8, 2.2]$ \cite{Cheng:2004ru};
with this values, the allowed range for $\Lambda$ is given by: $2.1 < \Lambda < 2.5\ GeV$.

\begin{figure}[]
\includegraphics[width=8cm]{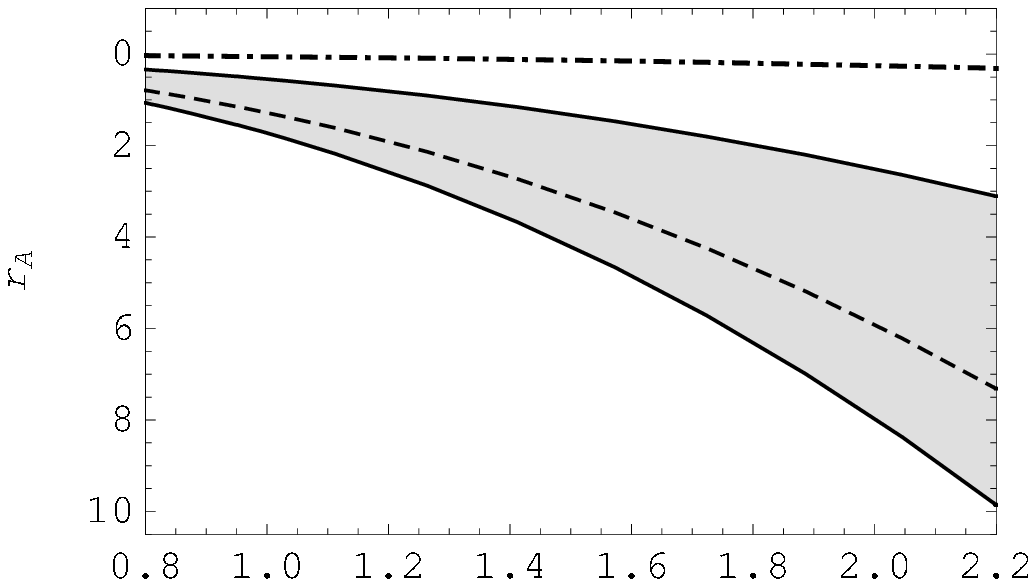}
\hspace{1truecm}
\includegraphics[width=8cm]{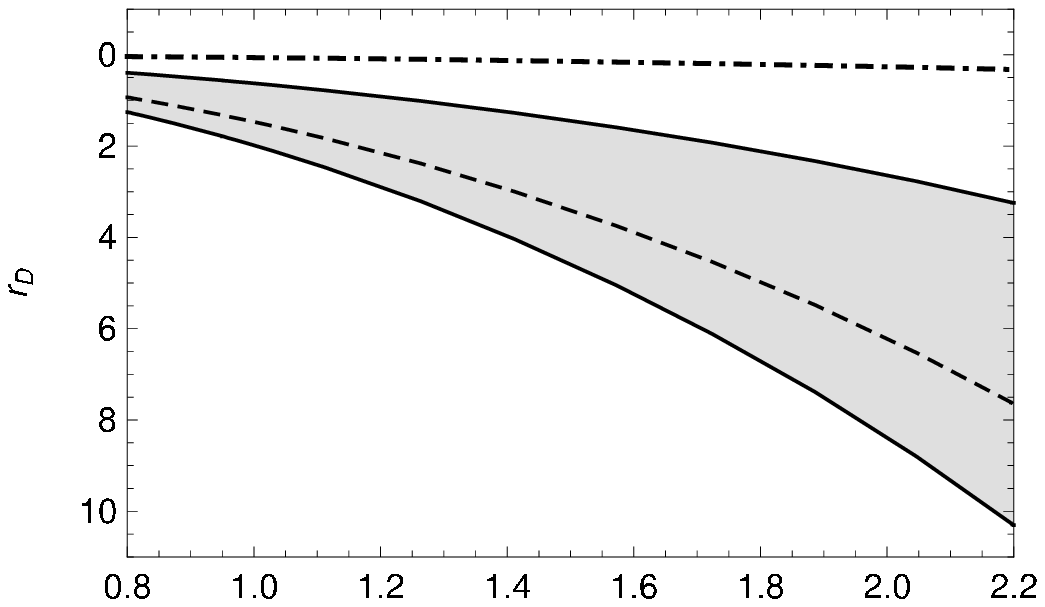}
\caption{The contributions coming from the loop graphs (for definitions see text).
The contributions are plotted for $g_{\eta_b
DD^\ast}/g_{\eta_bJJ }\approx $ 1 (dashed-dotted lines) and $g_{\eta_b
DD^\ast}/g_{\eta_bJJ }\approx \{11, 35\}$ (solid lines).
The dashed lines correspond to $g_{\eta_b DD^\ast}/g_{\eta_bJJ }\approx 26$.}
\label{f:rArD}
\end{figure}

In figure \ref{f:rArD}, left panel (right panel) the ratio
$r_{A} = A_{LD}\  g_{\eta _b DD^* }/g_{\eta _b JJ }$
($r_{D} = D_{LD}\  g_{\eta _b DD^* }/g_{\eta _b JJ }$)
is plotted as a function of $\alpha$ for the allowed value and the range of couplings ratio.
Moreover, the dashed lines are for $ g_{\eta _b DD^* }/g_{\eta _b JJ }\approx 26 $
which correspond to the central value in the allowed range for
$\eta_b \to D \overline{D^\ast}$ estimated in Ref. \cite{Maltoni:2004hv}.
It is clear that for  $ g_{\eta _b DD^* }/g_{\eta _b JJ }\approx 1$ the effects of the final
state interactions are negligible independently of $\alpha$.

Very different is the case in which the annihilation of $\eta_{b}$
into $D\overline{D^{\ast}}$ is large \cite{Maltoni:2004hv}.
The effects of final-state interactions could be large and
depend strongly on the value of $\alpha$ (cfr gray bands in
figure \ref{f:rArD}).

Starting from the estimate of the short-distance part in \cite{Jia:2006rx}
we are able to give the allowed range for the full branching ratio
\begin{equation}
{\cal B}r[\eta_b\to J/\psi\ J/\psi] = 0.5 \times 10^{-8} \div 1.2\times 10^{-5},
\label{e:finalris}
\end{equation}
where the lower bound corresponds to the corresponding one in \cite{Jia:2006rx}, while
the upper bound is obtained using the upper value in \cite{Jia:2006rx} and for
$\alpha = 2.2$, $g_{\eta_b DD^\ast}/g_{\eta_bJJ }= 35$.
The wide range for $\mathcal{B}r[\eta_b\to J/\psi\ J/\psi]$ in Eq. (\ref{e:finalris})
depends on the large theoretical uncertainty of the estimate of $\mathcal{B}r[\eta_b \to D \overline{D^\ast}]$ and
on the dependence on $\alpha$ parameter. It should be observed that
in \cite{Cheng:2004ru} the preferred value for $\alpha$ is $\alpha \approx 2.2$
for diagrams with $D$ and $D^\ast$ in $t$-channel, whereas a direct
calculation or measurement of the $\eta_b \to D \overline{D^\ast}$ process is in
order.

Finally we give an estimate of the discovery potential of the decay mode in
the LHC experiments. Each $J/\psi$ in the final state can be reconstructed
by means of its muonic decay mode which represents about 6\% of the total width,
so we have ${\cal B}r[\eta_b\to J/\psi\ J/\psi\to 4 \mu] \approx 2 \times 10^{-11} \div 4 \times 10^{-8}$.
Moreover, assuming, as in \cite{Jia:2006rx}, that i) the $\eta_b$ production cross section at LHC is
about 15 ${\rm \mu b}$ and ii) the integrated luminosity (per year) is about 300 ${\rm fb}^{-1}$,
the theoretically expected events are between $80$ and $2\times 10^5$. Experimentally
we have to consider also the product of acceptance and efficiency for detecting
$J/\psi$ decay to $\mu^+\mu^-$ which is of the order of $0.1$ \cite{Braaten:2000cm}, so
we expect between 0.8 and 2000 observed events per year.
Further, if we loose the constraint that $J/\psi$ must be tagged by
$\mu^+\mu^-$ pair and also allow its reconstruction through
$e^+e^-$ mode, we can have 3 $\div$ 8000 observed 4-lepton events per year.
These results seem to indicate that the chance of observing $\eta_b$ at LHC through the 4-lepton mode exists.

\begin{acknowledgments}

\noindent
I thank G.  De Nardo and D. Monorchio for a discussion on the experimental results from \textsc{Babar}.
\end{acknowledgments}


\end{document}